\documentclass[reprint,amsmath,amssymb,aps,prl,superscriptaddress]{revtex4-2}
\usepackage{array}
\usepackage{amsmath}
\usepackage{bm}
\usepackage{color}
\usepackage{float}
\usepackage{graphicx}
\usepackage{hyperref}
\hypersetup{colorlinks=true,allcolors=blue}
\usepackage{natbib}
\usepackage{newtxtext}
\usepackage{newtxmath}
\usepackage[caption=false]{subfig}
\newcommand\numberthis{\addtocounter{equation}{1}\tag{\theequation}}

\begin{document}

\title{Dynamical Signatures of Rank-2 $U(1)$ Spin Liquids}

\author{Emily Z. Zhang}
\affiliation{Department of Physics, University of Toronto, Toronto, Ontario M5S 1A7, Canada}

\author{Finn Lasse Buessen}
\affiliation{Department of Physics, University of Toronto, Toronto, Ontario M5S 1A7, Canada}

\author{Yong Baek Kim}
\affiliation{Department of Physics, University of Toronto, Toronto, Ontario M5S 1A7, Canada}
%\email[]{}
%\thanks{}

%\date{\today}

\begin{abstract}
Emergent $U(1)$ gauge theories and artificial photons in frustrated magnets are outstanding examples of many-body collective phenomena. The classical and quantum regimes of these systems provide platforms for classical and quantum spin liquids, and are the subject of current active theoretical and experimental investigations. Recently, realizations of rank-2 $U(1)$ (R2-U1) gauge theories in three-dimensional frustrated magnets have been proposed. Such systems in the quantum regime may lead to the so-called fracton ordered phases -- a new class of topological order that has been associated with quantum stabilizer codes and holography. However, there exist few distinguishing characteristics of these states for their detection in real materials. Here we focus on the classical limit, and present the dynamical spin structure factor for a R2-U1 spin liquid state on a breathing pyrochlore lattice. Remarkably, we find unique signatures of the R2-U1 state, and we contrast them with the results obtained from a more conventional $U(1)$ spin liquid. These results provide a new path of investigation for future inelastic neutron scattering experiments on candidate materials. 
\end{abstract}

%\keywords{}

\maketitle


\textit{Introduction.}--- Brought forth from magnetic materials with frustrated spin interactions is a rich world of unconventional phases and emergent laws of nature. Quantum spin liquids (QSL) -- one of the most exotic states known to condensed matter physics thus far -- evade ordering and remain highly entangled even down to zero temperature\cite{Savary2017a, Witczak-Krempa2014, Zhou2017a, Broholm2020}. These highly frustrated states can lead to remarkable properties, such as fractionalized quasiparticle excitations and emergent gauge theories, thus their detection in real materials has garnered both fundamental and practical interests. \

A well-studied example of a QSL state is the quantum spin ice, whose low energy excitations on the pyrochlore lattice gives rise to $U(1)$ electrodynamics with emergent photons and magnetic monopoles\cite{Hermele2004a, Banerjee2008, Castelnovo2008, Ross2011, Benton2012, Shannon2012, Savary2012, Kimura2013, Hao2014, Gingras2014, Kato2015, Chen2017, Savary2017b, Huang2018, Gao2019, Gaudet2019, Bramwell2020}. Another novel branch of QSLs in three dimensions has been suggested for systems that effectively mimic higher rank electrodynamics\cite{Xu2006, Benton2016, Pretko2017a, Pretko2017, Yan2020a,Yan2021}, wherein rank-2 or higher electromagnetic tensor fields can emerge out of the strongly interacting spins. Higher-rank tensor gauge theories in quantum systems can give rise to a new type of topological order, known as fractonic phases\cite{Vijay2016, Benton2016, Seiberg2020, Han2021}. Fractons, the charged excitations of these higher rank systems, have been associated with unusual behaviour, such as restricted motion in space\cite{Chamon2005, Haah2011a, Vijay2015, Vijay2016, Pretko2017a} and mimicking gravity\cite{Xu2006,Benton2016,Pretko2017d}. Applications of fractonic phases have been proposed in quantum local stabilizer codes\cite{Haah2011a,Schmitz2018, Kubica2018} and holography\cite{Yan2019}. Hence, both the quantum and classical regimes of such theories are of great general interest.\

\begin{figure}
\includegraphics[scale=0.5]{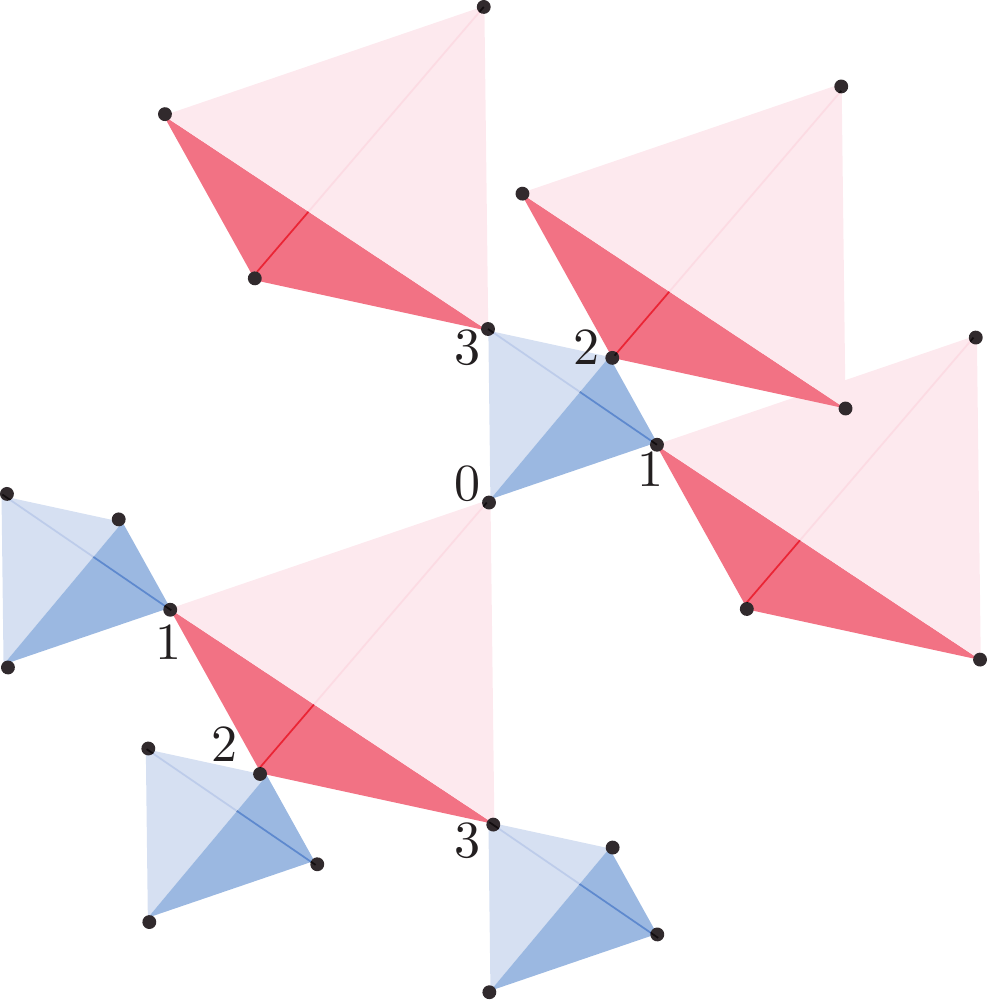}
\caption{\label{bp} The breathing pyrochlore lattice with $A$-tetrahedra (blue) and $B$-tetrahedra (red) of different sizes. The fluctuations of the low-energy modes of the $A$-tetrahedra are influenced by the surrounding $B$-tetrahedra. The spin moments are situated on the vertices of the corner-sharing tetrahedra, labelled by $0, 1, 2, 3$.}
\end{figure}

Recent studies of a classical spin model\cite{Yan2020a} have shown that highly frustrated spins situated on a breathing pyrochlore lattice (Fig. \ref{bp}) behave like an electric field with rank-2 tensor character, earning this state the name of a rank-2 $U(1)$ (R2-U1) spin liquid. At finite temperature, there exists a regime in which the R2-U1 state is stable. In this region, four-fold pinch point (4FPP) singularities in the equal-time spin correlations on certain momentum planes were observed\cite{Yan2020a,Prem2018}, which are distinct from the usual two-fold pinch point singularities found in spin ice\cite{Isakov2004, Benton2012}. The four-fold structures were seen in the spin flip channel of the structure factor, and are a preliminary distinguishing characteristic of the R2-U1 state\cite{Yan2020a}. Whether there are unique dynamic signatures of the R2-U1 state measurable in inelastic neutron scattering experiments remains to be investigated, and serves as the motivation for our studies. \

In this letter, we demonstrate novel behaviour in the inelastic spin structure factor of the R2-U1 state in the classical limit. We consider a spin model on a breathing pyrochlore lattice (Fig. \ref{bp}) with Heisenberg antiferromagnetic (HAF) exchange and Dzyaloshinskii-Moriya (DM) antisymmetric exchange interactions. Using classical finite temperature Monte Carlo techniques and molecular dynamics, we demonstrate that the 4FPP persists in the dynamic structure factor at low energies. Most notably, we contrast the R2-U1 signatures with those from the usual classical $U(1)\times U(1)\times U(1)$ Heisenberg spin liquid, which we refer to as the $U(1)$ spin liquid from this point forward for brevity. We demonstrate that the signatures of these two frustrated states are highly distinct from one another. We then discuss the potential connections of these classical results to the dynamics of frustrated quantum spin systems. Our results provide an important avenue of exploration for future inelastic neutron scattering experiments on material candidates for the R2-U1 spin liquid.

\textit{Model.}---
We consider a HAF model with DM interactions on the breathing pyrochlore lattice, composed of spins residing on the vertices of two different corner-sharing tetrahedra $A$ and $B$. The microscopic Hamiltonian is described by 
\begin{align*}
	\mathcal{H} &= \sum_{ij\in A} \left[ J_A\mathbf{S}_i \cdot \mathbf{S}_j 
	+ D_A\hat{\mathbf{d}}_{ij}\cdot(\mathbf{S}_i\times\mathbf{S}_j)\right] \\
	&+ \sum_{ij\in B} \left[ J_B\mathbf{S}_i \cdot \mathbf{S}_j 
	+ D_B\hat{\mathbf{d}}_{ij}\cdot(\mathbf{S}_i\times\mathbf{S}_j)\right],\numberthis \label{hamiltonian}
\end{align*}
where the bond-dependent vectors $\hat{\mathbf{d}}_{ij}$ can be found in \citep{SM}. Treating the spins classically, we can write this model in terms of the coarse-grained fields $\mathbf{m}_\Gamma$ that transform according to the irreducible representations $\Gamma=\{\mathsf{A}_2, \mathsf{E}, \mathsf{T}_2, \mathsf{T}_{1+}, \mathsf{T}_{1-} \}$ of the tetrahedra point group $\mathsf{T}_d$\cite{Yan2013}. The Hamiltonian then becomes 

\begin{align}
	\mathcal{H} &= \frac{1}{2}\sum_{\Gamma}a_{\Gamma,A}m^2_{\Gamma,A} + 
				\frac{1}{2}\sum_{\Gamma}a_{\Gamma,B}m^2_{\Gamma,B},
\end{align}
with the fields $\mathbf{m}_\Gamma$ and couplings $a_{\Gamma}$ defined in Refs. \cite{Yan2013, Yan2020a, Han2021}. $a_{\Gamma,A}$ and $a_{\Gamma_B}$ are functions of the spin exchange interactions, and their explicit forms can be found in \citep{SM}. We are interested in a minimal parameter choice in which the R2-U1 state has been shown to arise, given by $J_A$, $J_B>0$, $D_A<0$, and $D_B=0$\cite{Yan2020a}.  For the $A$-tetrahedra, this leads to the hierarchy 
\begin{align}
	a_{\mathsf{E},A}=a_{\mathsf{T}_{1-,A}}<a_{\mathsf{A}_2,A},a_{\mathsf{T}_{2,A}},a_{\mathsf{T}_{1+,A}}
\end{align}
while on the $B$-tetrahedra, we have 
\begin{align}
	a_{\mathsf{E},B}=a_{\mathsf{T}_{1-,B}}=a_{\mathsf{A}_2,B}=a_{\mathsf{T}_{2,B}}<a_{\mathsf{T}_{1+,B}}.
\end{align}
These hierarchies dictate the low energy physics of the system. For example, it is energetically costly for the $\mathsf{T}_{1+,B}$ mode to fluctuate, hence we set $\mathbf{m}_{\mathsf{T}_{1+,B}}=0$. In the low energy limit, $\mathbf{m}_\mathsf{E}$ and $\mathbf{m}_{\mathsf{T}_{1-}}$ can fluctuate on the $A$-tetrahedra, while the other fields must vanish. 

The condition $\mathbf{m}_{B,\mathsf{T}_{1+}}=0$ on the $B$-tetrahedra imposes additional constraints on the $A$-tetrahedra\cite{Yan2020a, Han2021}. These constraints can be expressed as a Gauss' law constraint of the fluctuating fields living on the $A$-tetrahedra, in the form of 
\begin{align}
	\partial_i E^{ij}_A =0. \label{gauss}
\end{align}
Using this constraint, a rank-2 electric field tensor $\mathbf{E}_A$ can be constructed in terms of the coarse-grained fields as 
\begin{align}
	\mathbf{E}_A = 
	\begin{pmatrix}
		\frac{2}{\sqrt{3}} m^1_{\mathsf{E},A}   &  m^z_{\mathsf{T}_{1-,A}}  &  m^y_{\mathsf{T}_{1-,A}} \\
		m^z_{\mathsf{T}_{1-,A}} & -\frac{1}{\sqrt{3}} m^1_{\mathsf{E},A}+m^2_{\mathsf{E},A} & m^x_{\mathsf{T}_{1-,A}}\\
		m^y_{\mathsf{T}_{1-,A}} & m^x_{\mathsf{T}_{1-,A}} & -\frac{1}{\sqrt{3}} m^1_{\mathsf{E},A}-m^2_{\mathsf{E},A} 
	\end{pmatrix}
\end{align}
with properties $E_{A}^{ij}=E_{A}^{ji}$ and $\text{Tr}\ \mathbf{E}_A=0$. These properties fit within the general framework of a self-dual, vector-charged, traceless form of R2-U1 electrodynamics \cite{Pretko2017a, Pretko2017} that obeys a Gauss' law described by Eq. \eqref{gauss} in the low energy limit. The four-fold pinch point (4FPP) singularity has been observed in the $hk0$ plane of the correlation function, which for a vector-charged traceless R2-U1 theory, has the form\cite{Prem2018}
\begin{align*}
	\left<E_{ij}(\mathbf{q})E_{kl}(\mathbf{-q})\right> & \propto  \frac{1}{2}(\delta_{ij}\delta_{jl}+\delta_{il} \delta_{jk}) + \frac{q_i q_j q_k q_l}{q^4}\\
	& -\frac{1}{2}\left(\delta_{ik}\frac{q_j q_l}{q^2} + \delta_{jk}\frac{q_i q_l}{q^2} + \delta_{il}\frac{q_j q_k}{q^2} +  \delta_{jl}\frac{q_i q_k}{q^2}\right)\\
	& -\frac{1}{2}\left(\delta_{ij} - \frac{q_i q_j} {q^2}\right) \left(\delta_{kl} - \frac{q_k q_l}{q^2}\right). \numberthis
\end{align*}
The 4FPP singularity appearing in the $E_{ij}$ correlation function can also be seen in the spin-flip channel of the equal-time spin structure factor in the $h0k$ and $hhk$ planes, which is natural since $E_{ij}$ are simply coarse-grained fields due to the spins.

In order to capture the dynamics of the system, we calculate the momentum- and energy-dependent spin correlation, i.e. the inelastic spin structure factor, given by 
\begin{align}
{S}^{\mu\nu}(\mathbf{q},\omega)=\frac{1}{2\pi N} \sum_{i,j=1}^N\int_{-\infty}^{\infty}\text{d}t\ e^{-i\mathbf{q}\cdot(\mathbf{r}_i-\mathbf{r}_j)+i\omega t} \left<S_{i}^{\mu}(t)S_{j}^{\nu}(0)\right>. 
\end{align}
We investigate the spectrum both with unpolarized neutrons
\begin{align}
	\mathcal{S}(\mathbf{q},\omega) = \sum_{\mu,\nu}\left( \delta_{\mu\nu} - \frac{q_\mu q_\nu}{q^2}\right)\mathcal{S}^{\mu\nu}(\mathbf{q},\omega),
\end{align}
and in the polarized spin-flip channel 
\begin{align}
	\mathcal{S}_{\text{SF}}(\mathbf{q},\omega) = \sum_{\mu,\nu}\left(v_\perp^\mu v_\perp^{\nu} \right)\mathcal{S}^{\mu\nu}(\mathbf{q},\omega),
\end{align}
with neutrons polarized in the $\hat{\mathbf{v}}$ direction perpendicular to the scattering channel, and $\hat{\mathbf{v}}_\perp=\frac{\hat{\mathbf{v}}\times\mathbf{q}}{|\hat{\mathbf{v}}\times\mathbf{q}|}$. Further details, along with additional information about the equal-time spin structure factor, can be found in \citep{SM}.

\textit{Numerical methods.}---
To study the finite temperature dynamics of the model in Eq. \eqref{hamiltonian}, we utilized finite temperature Monte Carlo (MC) techniques\cite{Moessner1998, Zhang2019}. We fixed the magnitude of the classical spins $\mathbf{S}=(S_x, S_y, S_z)$ to be $S=1/2$. We studied system sizes of $4\times L \times L \times L$, up to a maximum of $L=20$. We allowed the system to thermalize using a combination of simulated annealing and parallel tempering -- both of which were done for at least $5\times10^5$ MC sweeps. After thermalization, we performed another $10^6$ MC steps with measurements recorded every $10$ steps.  

According to the phase diagram shown in Ref. \cite{Yan2020a}, the R2-U1 state is stable within a finite temperature window between $T_1\approx 4\times10^{-3}|J_A|$ and $T_2\approx  5\times 10^{-2}|J_A|$ for the parameter set $(J_A,J_B,D_A,D_B)=(1,1,-0.15,0)$. Above $T_2$, there is a cross-over from the R2-U1 spin liquid state to the $U(1)$ spin liquid. The two phases can be distinguished from each other by their pinch point characteristics in the equal-time structure factor, as shown in Ref. \cite{Yan2020a}. More specifically, there is an evolution of the pinch point from a four-fold singularity to a two-fold one at this temperature. Within the R2-U1 temperature window, we used the many degenerate spin configurations obtained from our MC simulations as the initial conditions (IC) for the molecular dynamics (MD) at various temperatures. We time-evolved each IC according to the semi-classical Landau Lifshitz equations of motion\cite{Moessner1998, Zhang2019}, and we numerically integrated and averaged over the configurations to obtain the dynamical spin structure factor. Details of the numerical methods used can be found in \citep{SM}. 

\begin{figure}
\includegraphics[scale=0.8]{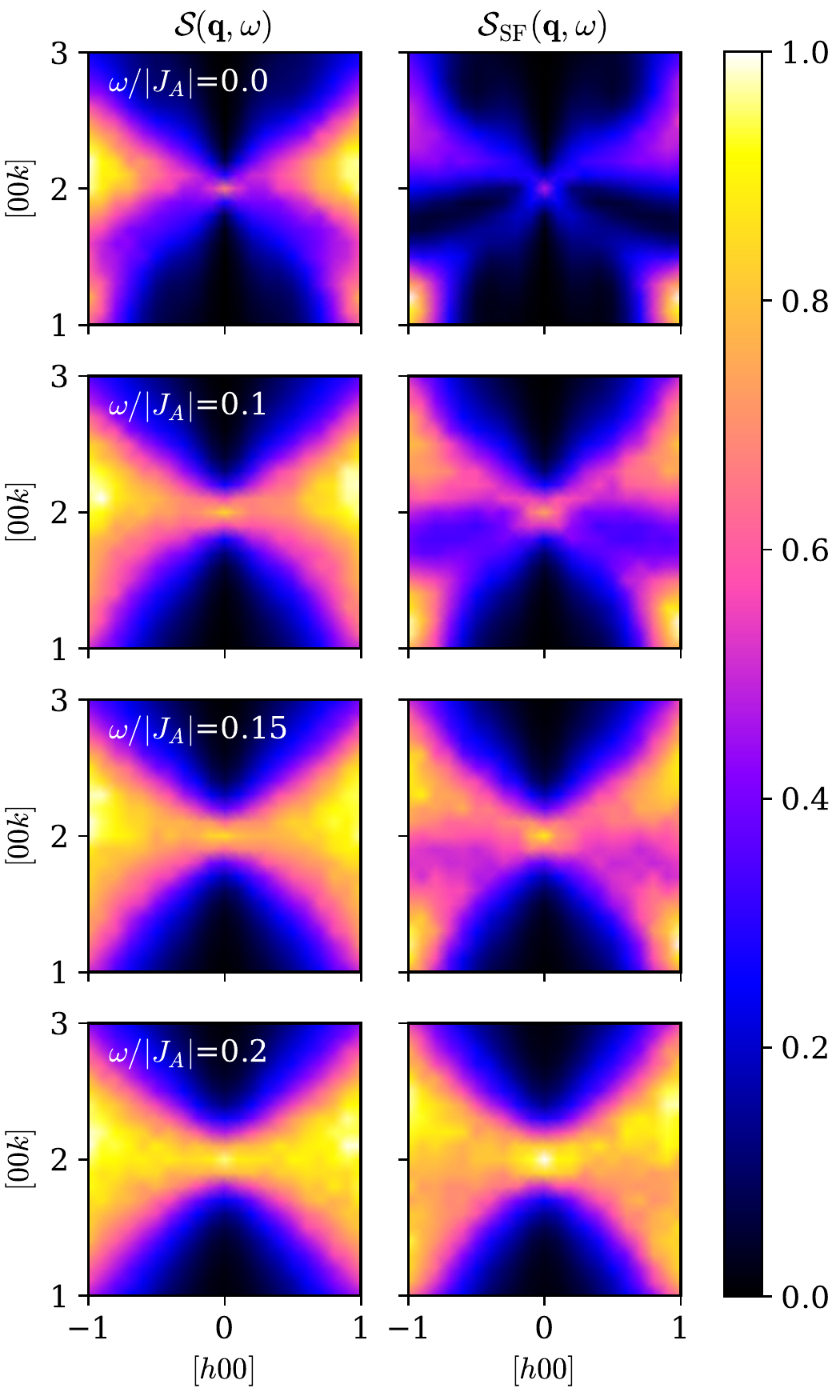}
\caption{\label{h0kplane} Momentum-dependent dynamical structure factor in the $[h0k]$ plane at varying energies, for the parameterization $(J_A,J_B,D_A,D_B)=(1,1,-0.15,0)$ at $T/|J_A|=0.01$. The four-fold pinch point structure is shown in the spin-flip (SF) channel in the right column, while the total dynamic structure factor is shown on the left column. The intensity scale for each panel has been normalized to arbitrary units. } 
\end{figure}

\begin{figure}
\includegraphics[scale=0.8]{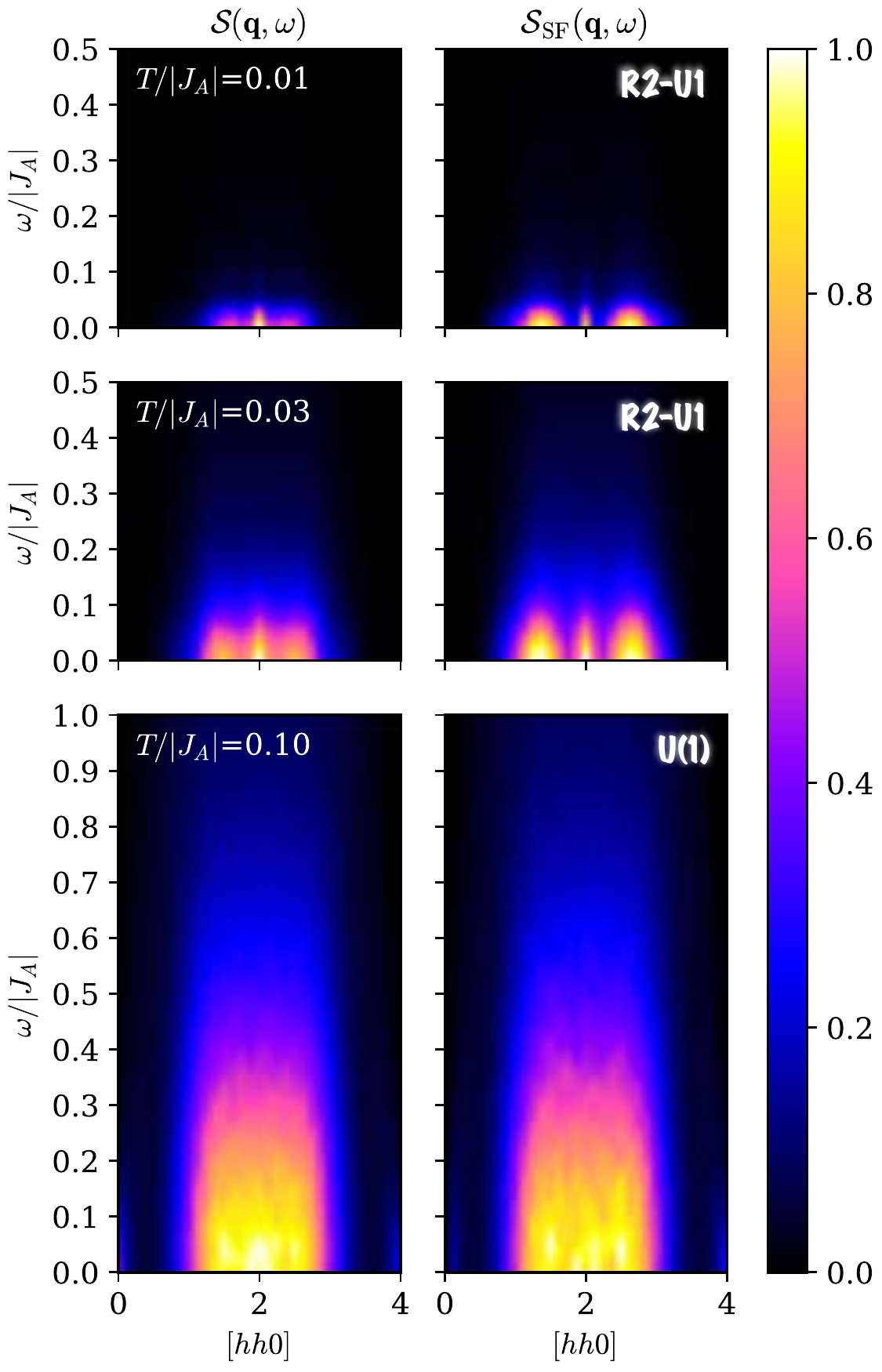}
\caption{\label{hh0} Energy dependence of dynamical structure factor along the $[hh0]$ momentum cut, for the parameterization $(J_A,J_B,D_A,D_B)=(1,1,-0.15,0)$. The total dynamic structure factor  (left column) and dynamic structure factor in the spin-flip (SF) channel (right column) are shown for $T/|J_A|=0.01$ and $T/|J_A|=0.03$ within the R2-U1 regime, and $T/|J_A|=0.10$ within the $U(1)$ spin liquid regime. The intensity scale for each panel has been normalized to arbitrary units.  }
\end{figure}

\begin{figure*}[t!]
\includegraphics[scale=0.8]{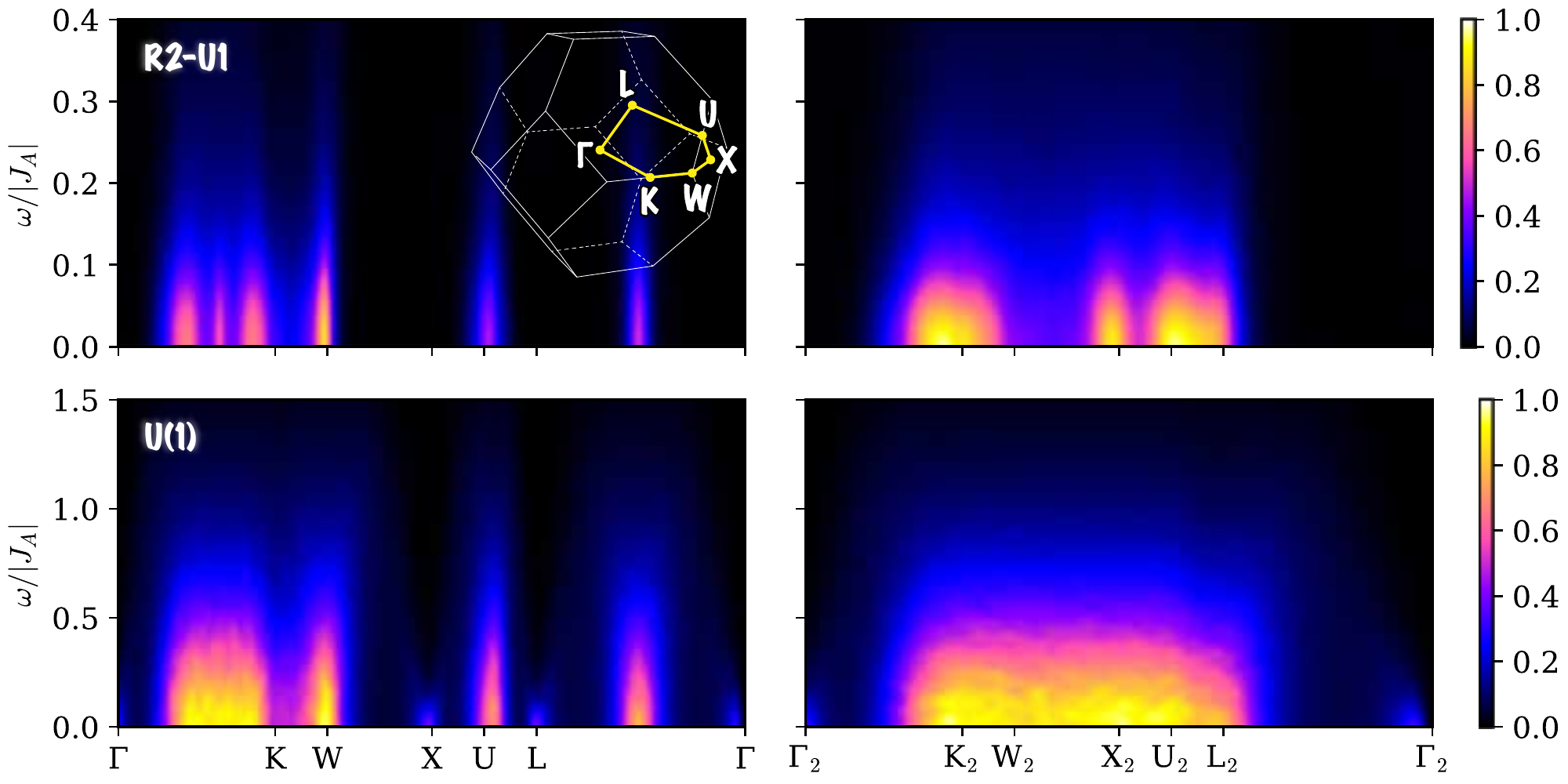}
\caption{\label{hsp} Energy dependence of dynamical structure factor in the spin flip channel along high symmetry directions in the first Brillouin zone (FBZ). The dynamical structure factor is plotted along the FBZ edge (left column) and inside the FBZ (right column), where $\Gamma_2$, $K_2$, $W_2$, $X_2$, $U_2$, and $L_2$ are shorthands for $1/2$ of each high symmetry point respectively. The temperatures used were $T/|J_A|=0.03$ within the R2-U1 regime, and $T/|J_A|=0.13$ within the $U(1)$ spin liquid regime. Note that the energy scales between the R2-U1 and $U(1)$ case differ. The intensity scale for each panel has been normalized to arbitrary units. }
\end{figure*}

\textit{Results.}--- The dynamical structure factor is plotted at varying energies as a function of momentum at fixed temperature, shown in Fig. \ref{h0kplane}, as a function of energy and momentum at increasing temperatures, shown in Fig. \ref{hh0}, and along high symmetry directions in momentum space, shown in Fig. \ref{hsp}.

Fig. \ref{h0kplane} depicts the normalized total and spin-flip dynamical structure factor in the $h0k$ plane at energies $0,0.1,0.15,$ and $0.2\ |J_A|$ for $T=0.01\ |J_A|$ where the R2-U1 state is stable. The 4FPP characteristic present in the equal-time structure factor is also seen in the static structure factor $\mathcal{S}_{\text{SF}}(\mathbf{q},\omega)$ in the spin-flip channel. As the energy increases, the four-fold nature of the pinch point gradually washes out to a two-fold pinch point characteristic of $U(1)$ spin liquids. Interestingly, the energy at which this cross-over occurs, i.e. approximately $0.2\ |J_A|$, is similar to the temperature above which the 4FPP disappears in the equal-time structure factor. 


Next, we present the energy dependence of the dynamical structure factor along the $hh0$ momentum path in Fig. \ref{hh0}. The first two rows are depicted for temperatures well within the R2-U1 regime, while the last row shows the dynamics in the $U(1)$ spin liquid regime. There are two important observations to note. First, as the temperature is increased from $0.01\ |J_A|$ to $0.03\ |J_A|$, the energy scale for the excitations also increases. Secondly, the triple-peak structure of the signatures in the R2-U1 state are easily differentiated from the broad structures seen in the $U(1)$ spin liquid case. 

Fig. \ref{hsp} shows the dynamical structure factor in the spin flip channel along a high symmetry path $\Gamma$-$K$-$W$-$X$-$U$-$L$-$\Gamma$ in the extended first Brillouin zone (FBZ), and along a path $\Gamma_2$-$K_2$-$W_2$-$X_2$-$U_2$-$L_2$-$\Gamma_2$ inside the extended FBZ. There are a few notable distinguishing features of the R2-U1 state to contrast with the $U(1)$ state. Along the FBZ edges, the multi-peak structure present in Fig. \ref{hh0} also appears along the $\Gamma$-$K$ path in the R2-U1 regime, whereas the signal is simply broad along this cut for the $U(1)$ spin liquid. On the second path within the FBZ, there is a stark contrast between the R2-U1 and $U(1)$ inelastic structure factors. The signal for the R2-U1 is suppressed between the $W_2$ and $X_2$ points, and between the $X_2$ and $U_2$ points, as opposed to the broad signal along these points in the $U(1)$ case. In other words, the signal supression in the inelastic structure factor along this second momentum path is another characteristic of the R2-U1 state, in addition to the 4FPP structure in seen in Fig. \ref{h0kplane}. 

\textit{Discussion.}---
In this letter, we demonstrated distinguishing characteristics of the R2-U1 state in its inelastic spin structure factor. Not only does the 4FPP characteristic persist at low energies, shown in Fig. \ref{h0kplane}, but both Figs. \ref{hh0} and \ref{hsp} demonstrate how the energy dependence of the dynamical structure factor can be used to distinguish the R2-U1 from the $U(1)$ state. Thus, our results illuminate a new path for experimentally detecting the R2-U1 state from inelastic neutron scattering in real materials.

Naturally, the question of whether these features would also be present in the quantum spin system arises. Solving the quantum model and its dynamics for 3D systems, however, has been a historically difficult feat. Meanwhile, classical simulations of frustrated spin systems using molecular dynamics have indicated good qualitative agreement with their quantum counterparts, even in the quantum spin liquid and quantum paramagnetic regimes\cite{Samarakoon2017, Samarakoon2018}. This type of semi-classical modeling has been done for Kitaev-like frustrated magnets, involving bond-dependent Kitaev\cite{Samarakoon2017} and off-diagonal $\Gamma$\cite{Samarakoon2018} interactions on a honeycomb. These studies showed that qualitative features seen in the classical inelastic spin structure factor persisted in the dynamics for the quantum system, implying that the highly degenerate classical states are participating in quantum fluctuations down to low energy scales. Due to this quantum-classical correspondence in the dynamics, we believe that our classical results provide invaluable insight into the putative R2-U1 quantum state. This work therefore serves as a reference point for future finite-temperature dynamical simulations for the quantum rank-2 $U(1)$ spin liquid state. Moreover, if the magnitude of the spin magnetic moments in real breathing pyrochlore materials are large, then our classical results would directly apply. Yb-based pyrochlore oxides such as Ba$_3$Yb$_2$Zn$_5$O$_{11}$\cite{Kimura2014, Haku2016, Rau2016, Rau2018}, whose Yb atoms form a breathing pyrochlore lattice, have been proposed as candidates to host the R2-U1 state, thus they are a potential playground for future inelastic neutron scattering experiments.

\begin{acknowledgements}
We thank SangEun Han for helpful discussions. We acknowledge support from the Natural Sciences and Engineering Research Council of Canada (NSERC) and the Centre of Quantum Materials at the University of Toronto. E.Z.Z. was further supported by the NSERC Canada Graduate Scholarships - Doctoral (CGS-D). The computations were performed on the Cedar cluster, which is hosted by WestGrid and SciNet in partnership with Compute Canada. 
\end{acknowledgements}

\bibliography{r2u1}

\end{document}

\title{Supplemental Material}

\author{Emily Z. Zhang}
\affiliation{Department of Physics, University of Toronto, Toronto, Ontario M5S 1A7, Canada}

\author{Finn Lasse Buessen}
\affiliation{Department of Physics, University of Toronto, Toronto, Ontario M5S 1A7, Canada}

\author{Yong Baek Kim}
\affiliation{Department of Physics, University of Toronto, Toronto, Ontario M5S 1A7, Canada}

\maketitle

\setcounter{equation}{0}
\setcounter{figure}{0}
\setcounter{table}{0}

\renewcommand{\thesection}{S\arabic{section}}
\renewcommand{\theequation}{S\arabic{equation}}
\renewcommand{\thefigure}{S\arabic{figure}}
\renewcommand{\thetable}{S\arabic{table}}

\onecolumngrid

\section{Details of the Coupling Constants}
As described in the main text, the model used was a breathing pyrochlore lattice with antiferromagnetic Heisenberg interactions and Dzyaloshinskii-Moriya (DM) interactions. In our work, we choose the parameter regime $J_A>0$, $J_B>0$, $D_A<0$, and $D_B=0$. Under this parameter choice, the couplings become 
\begin{align}
	a_{\mathsf{E},A} &= a_{\mathsf{T}_{1-,A}} = -J_A + 2D_A/\sqrt{2} \\
	a_{\mathsf{A}_2,A} &= -J_A - 4D_A/\sqrt{2} \\
	a_{\mathsf{T}_{2,A}} &= -J_A-2D_A/\sqrt{2} \\ 
	a_{\mathsf{T}_{1+,A}} &= 3J_A
\end{align}
on the $A$-tetrahedra and 
\begin{align}
	a_{\mathsf{E},B}&=a_{\mathsf{T}_{1-,B}}=a_{\mathsf{A}_2,B}=a_{\mathsf{T}_{2,B}}=-J_B\\
	a_{\mathsf{T}_{1+,B}}&=3J_B
\end{align}
on the $B$-tetrahedra. On the $A$-terahedra, because of the parameter choice, $a_{\mathsf{E},A} = a_{\mathsf{T}_{1-,A}}$ will always be negative, while the other constants remain positive. Similarly, $a_{\mathsf{E},B}=a_{\mathsf{T}_{1-,B}}=a_{\mathsf{A}_2,B}=a_{\mathsf{T}_{2,B}}$ will always be negative, while $a_{\mathsf{T}_{1+,B}}$ is always positive. This observation informs the hierachies in Eqs. 3 and 4 in the main text.
 
\section{Details of the Numerical Methods}
\begin{figure}[h!]
	\begin{center}
	\includegraphics[scale=0.7]{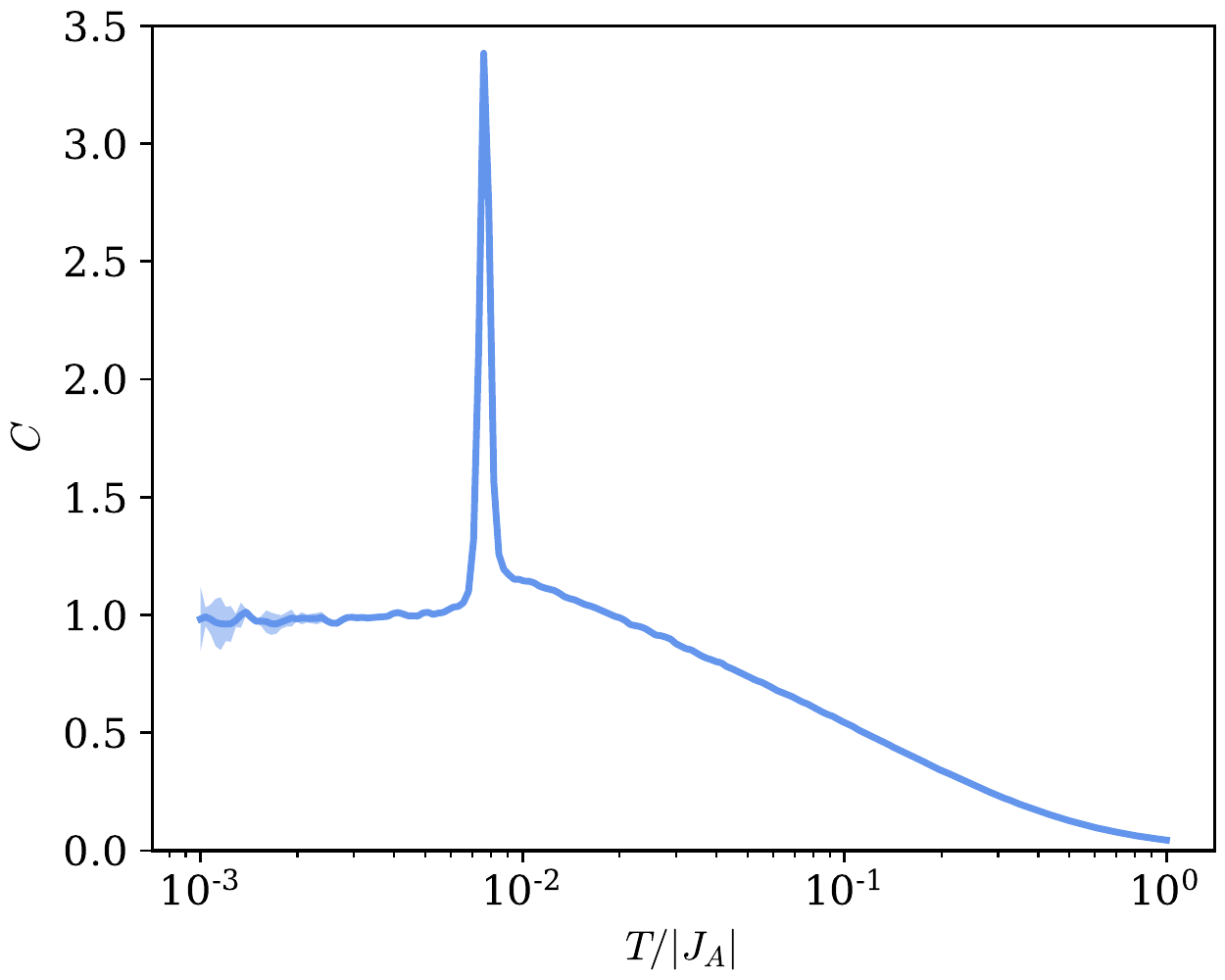}
	\caption{Heat capacity per spin $C$ as a function of temperature for $L=4$. The interaction parameters used were $(J_A,D_A,J_B,D_B)=(1,-0.15,1,0)$. The shaded region represents the error of the measurements. \label{cv} }
	\end{center}
\end{figure}
As mentioned in the main text, the Hamiltonian of the model is given by 
\begin{align}
	\mathcal{H} = \sum_{ij\in A} \left[ J_A\mathbf{S}_i \cdot \mathbf{S}_j 
	+ D_A\hat{\mathbf{d}}_{ij}\cdot(\mathbf{S}_i\times\mathbf{S}_j)\right] 
	+ \sum_{ij\in B} \left[ J_B\mathbf{S}_i \cdot \mathbf{S}_j 
	+ D_B\hat{\mathbf{d}}_{ij}\cdot(\mathbf{S}_i\times\mathbf{S}_j)\right], \label{ham}
\end{align}
where $ij\in A$ and $ij\in B$ describe the nearest neighbour bonds on the $A$- and $B$-tetrahedra respectively. The centres of the $A$-tetrahedra lie on the fcc lattice, whose basis vectors are given by $\mathbf{a}_1=\frac{1}{2}(0,1,1),\mathbf{a}_2=\frac{1}{2}(1,0,1)$, and $\mathbf{a}_3=\frac{1}{2}(1,1,0)$. The sites of each tetrahedra are enumerated as 0, 1, 2, 3 as shown in Fig. 1 in the main text. Their positions relative to the centres of the $A$-tetrahedra are given by $\mathbf{r}_0=\frac{a}{8}(1,1,1);\  \mathbf{r}_1=\frac{a}{8}(1,1,-1);\ \mathbf{r}_2=\frac{a}{8}(-1,1,-1)$, and $\mathbf{r}_3=\frac{a}{8}(-1,-1,1)$. The bond-dependent vectors $\hat{\mathbf{d}}_{ij}$ are defined as $\hat{\mathbf{d}}_{01} = \frac{1}{\sqrt{2}} (0,-1,1);\ \hat{\mathbf{d}}_{02} =\frac{1}{\sqrt{2}} (1,0,-1);\ \hat{\mathbf{d}}_{03}=\frac{1}{\sqrt{2}}(-1, 1, 0);\  \hat{\mathbf{d}}_{12} =\frac{1}{\sqrt{2}}(-1,-1,0);\ \hat{\mathbf{d}}_{13} =\frac{1}{\sqrt{2}}(1,0,1);\ $ and $\hat{\mathbf{d}}_{23} =\frac{1}{\sqrt{2}} (0,-1,-1)$ \cite{Kotov2005, Canals2008, Rau2016}.

\subsection{Monte Carlo}

We used Monte Carlo (MC) techniques to obtain the spin configurations for use in the molecular dynamics (MD) simulations. The spins were treated classically, i.e. $\mathbf{S}=(S_x, S_y, S_z)$, and their magnitudes were fixed to $S=1/2$. The MC simulations were executed in three stages: simulated annealing, thermalization sweeps, and measurement sweeps. Firstly, we performed simulated annealing from a high temperature to the desired temperature $T$. We then used a combination of parallel tempering and overrelaxation for the rest of the thermalization process. 10 overrelaxation steps were performed for each attempted metropolis update, and replica exchanges were attempted once every 10 metropolis updates. We ran the thermalization phase for at least $5\times 10^{5}$ Monte Carlo (MC) sweeps. The same process was used for the measurement phase, except the spins were outputted every 10 sweeps. There were at least $5\times 10^{6}$ measurement sweeps performed, and these measurements were used to compute the thermodynamic observables, the equal-time structure factors, and the dynamical spin structure factors. A total of 2000 initial configurations were used in the MD simulations.

Fig. \ref{cv} shows the heat capacity $C$ as a function of temperature for system size $L=4$. For this system size, the results are well-converged, and there is a clear transition from the so-called $q=W$ order to the R2-U1 phase at ~$T=8\times 10^{-2}|J_A|$. However, we note that for large system sizes, low-temperature heat capacity does not converge well in the $q=W$ regime. Regardless, this temperature range is outside of our region of interest, and further exploration of this ordered state is outside the scope of this work. We also note that there is no thermodynamic signature of a crossover between the R2-U1 and $U(1)$ spin-liquid regions. Thus, the only known distinguishing characteristics between these two phases are the equal-time structure factor, as presented by Ref. \cite{Yan2020a}, and the dynamical spin structure factors, presented in this work. 

\subsection{Equal-Time Structure Factor}
\begin{figure}[h!]
	\begin{center}
	\includegraphics[scale=0.6]{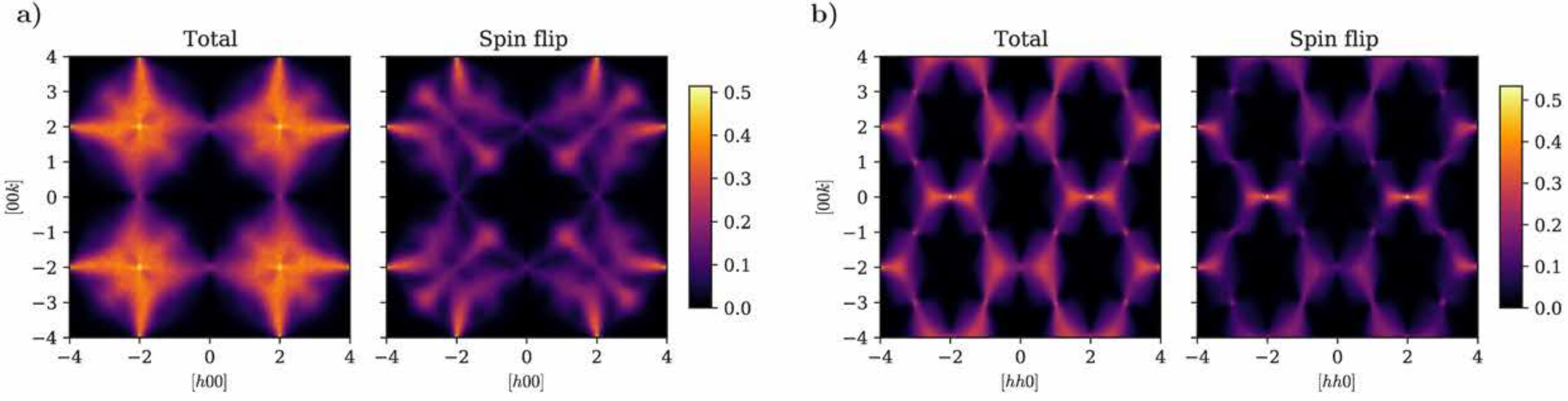}
	\caption{Equal-time structure factor as a function of momentum in a) the $h0k$ plane and b) in the $hhk$ plane. The results for the total and spin-flip channel of the structure factor are shown. \label{SSF} }
	\end{center}
\end{figure}

The equal-time structure factor is defined as
\begin{align}
	\left<S^\mu(\mathbf{q})S^\nu(\mathbf{-q})\right> = \frac{1}{N} \sum_{i, j}^N  \left<S^\mu(\mathbf{r}_i) S^\nu(\mathbf{r}_j)\right>e^{i\mathbf{q}\cdot(\mathbf{r}_i-\mathbf{r}_j)}
\end{align}
where $N$ is the number of sites, and $\mu,\nu$ label the spin components $x, y, z$. The total and spin-flip channels are defined the same way as Eqs. 9 and 10 in the main text. Fig. \ref{SSF} shows that the four-fold pinch points can be seen in at the $[200]$ and equivalent points in the spin-flip channel, as seen in Ref. \cite{Yan2020a}. As shown in the main text, the four-fold pinch point in the equal-time structure factor can also be seen in the static spin structure factor $\mathcal{S}(\mathbf{q},\omega=0)$. 

\subsection{Molecular Dynamics}
\begin{figure}[h!]
	\begin{center}
	\includegraphics[scale=0.8]{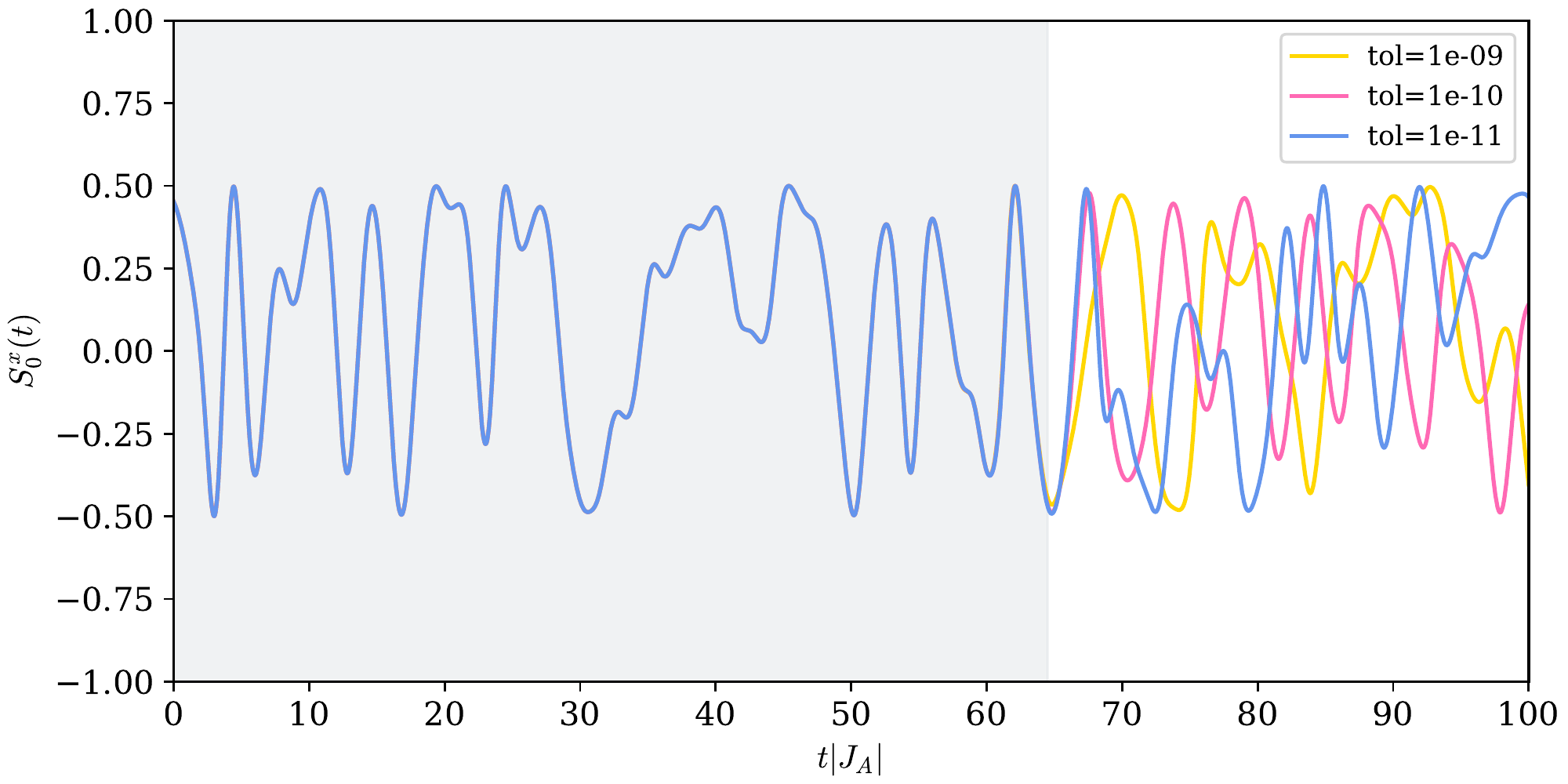}
	\caption{Time evolution of the $x$-component of spin as a function of time for an arbitrary lattice site for $L=4$. The results of the time evolution are shown for three error tolerances, and are converged within the grey area. \label{convergence} }
	\end{center}
\end{figure}

The time evolutions of the classical spins can be found by solving the Landau-Lifshitz-Gilbert (LLG) equations of motion\cite{Lakshmanan2011}, given by 
\begin{align}
\frac{d}{dt}\mathbf{S}_i=\{\mathbf{S}_i,\mathcal{H}\},
\end{align}
where $\{\mathcal{A},\mathcal{B}\}$ is the Poisson bracket between functions of spin $\mathcal{A}$ and $\mathcal{B}$. The results from the MC simulations are taken as the initial configurations for solving the LLG equations. To solve this system of coupled ordinary differential equations, we used a 7th order Runge Kutta method\cite{Rackauckas2017} with an error tolerance of $10^{-9}$. Fig. \ref{convergence} shows the convergence for different error tolerances using this ODE differential solver. The three tolerances are converged up to $t\simeq 65/|J_A|$ (grey area). For all our simulations, we chose a time window $t=60/|J_A|$, which is well within the converged area for the different error tolerances. This time window is also long enough to capture multiple periods of oscillation. 

\bibliography{r2u1}